\documentclass[12pt]{article}

\usepackage[cp1251]{inputenc}
\usepackage[russian]{babel}
\usepackage{graphicx,comment}
\usepackage{amsmath,amssymb,amsfonts,amsthm,comment}

\begin{document}

\noindent {\sl Problems of Information Transmission},\\
\noindent v. 53, no. 3, pp. 3--15, 2017.

\begin{center} {\bf M. V. Burnashev} \end{center}

%\vskip 0.4cm

\begin{center}
{\large\bf TWO THEOREMS ON DISTRIBUTION OF GAUSSIAN QUADRATIC
FORMS}
\footnote[1]{The research was carried out at the IITP RAS at the
expense of the Russian Foundation for Sciences
(project 14-50-00150).}
\end{center}

{\begin{quotation} \normalsize New results on comparison of
distributions of Gaussian quadratic forms are presented.
\end{quotation}}

\vskip 0.7cm

Let $\xi_{1},\ldots,\xi_{n}$ -- independent
${\mathcal N}(0,1)$-Gaussian random variables. For \\
$\boldsymbol{a} = (a_{1},\ldots,a_{n}) \in \mathbb{R}_{+}^{n}$ and
$x \geq 0$ consider the following probability
$$
\beta(x,\boldsymbol{a}) = \mathbf{P}
\left(\sum\limits_{i=1}^{n}a_{i}\xi_{i}^{2}< x\right).
$$

We are interested in what
$\boldsymbol{a}, \boldsymbol{b} \in \mathbb{R}^{n}_{+}$ and $x$
the following inequality holds
\begin{eqnarray}\label{Bak1}
\beta(x,\boldsymbol{a}) \leq \beta(x,\boldsymbol{b}).
\end{eqnarray}

Below the vector $\boldsymbol{a} \in \mathbb{R}_{+}^{n}$
is called {\it monotone}, if
$a_{1} \geq a_{2} \geq \ldots \geq a_{n} \geq 0$.
As usually, $\boldsymbol{a} \geq \boldsymbol{b}$ means
$a_{i} \geq b_{i}$, $i=1,\ldots,n$.

{\bf 1. Known comparison theorem}.
In \cite[Theorem 1]{Bak95} the following result was proved.
Let $\boldsymbol{a}, \boldsymbol{b} \in \mathbb{R}_{+}^{n}$ -
monotone vectors and the following condition for them is fulfilled
\begin{eqnarray}\label{Bak2}
\sum\limits_{i=1}^{k}a_{i} \geq \sum\limits_{i=1}^{k}b_{i},
\qquad k = 1,\ldots,n.
\end{eqnarray}
Then for any $x \geq 2\sum\limits_{i=1}^{n}b_{i}$
the inequality \eqref{Bak1} holds.

{\it Remark} 1. In order \eqref{Bak2} to be valid, we need, in
particular, $\max\limits_{i}a_{i} \geq \max\limits_{i}b_{i}$,
what is rather restrictive.

Inequality \eqref{Bak1} is useful in problems of detection of
stochastic signals in Gaussian noise. But application of Theorem 1
from \cite{Bak95} in such problems is rather difficult because of
the restrictive assumption \eqref{Bak2} and the requirement
$x \geq 2\sum\limits_{i=1}^{n}b_{i}$ (in stochastic signals
detection problems the inequality \eqref{Bak1} is usually
required for $x < \sum\limits_{i=1}^{n}b_{i}$).

In the paper the inequality \eqref{Bak1} is proved under different
from \eqref{Bak2} assumption. First, a simple similar
Proposition 1 is proved, and then it is strengthened using
additional arguments (Theorems 1 and 2).

Concerning applications of the inequality \eqref{Bak1} it should
be mentioned that such results are helpful in problems of detection
of Gaussian stochastic signals in the background of independent
additive Gaussian noise [2--5].
%\cite{Wald, Leman,Bur79, ZhangPoor11}.
Consider, for example, the problem of detection of Gaussian
stochastic signal vector ${\mathbf s}$ in the background of
independent additive Gaussian noise $\boldsymbol{\xi}$. If the
vector $\boldsymbol{\sigma}$ of the vector ${\mathbf s}$
intensities is known, then the logarithm of the corresponding
likelihood ratio in that problem is a Gaussian quadratic form,
similar to one considered above. Assume that we know only that
the vector $\boldsymbol{\sigma}$ belongs to the given set
${\mathcal E}$. Then natural question arises: is it possible to
replace the set ${\mathcal E}$ by a smaller set ${\mathcal E}_{0}$
without loss of detection quality (in particular, to replace
${\mathcal E}$ by a single point $\boldsymbol{\sigma}_{0}$) ?
Such problem will be considered by author in the paper
\cite{Bur17}.

Some results showing validity of the inequality \eqref{Bak1}
can also be found in \cite{Ponom85, Bak95}.

{\bf 2. The first result}. For vectors
$\boldsymbol{a},\boldsymbol{b} \in \mathbb{R}^{n}_{+}$
introduce functions
\begin{equation}\label{deff1}
\begin{gathered}
f(i,\boldsymbol{a},\boldsymbol{b}) = \frac{1}{b_{i}} -
\frac{1}{a_{i}}, \quad i = 1,\ldots,n, \qquad
D(\boldsymbol{a},\boldsymbol{b}) =
\left(\prod\limits_{i=1}^{n}(a_{i}/b_{i})\right)^{1/2}.
\end{gathered}
\end{equation}

P r o p o s i t i o n \,1. {\it Let
$\boldsymbol{a}, \boldsymbol{b} \in \mathbb{R}^{n}_{+}$ and
$D(\boldsymbol{a},\boldsymbol{b}) > 1$.
If $x$ satisfies the condition }
\begin{equation}\label{prop1a}
\begin{gathered}
x \leq \sup_{\substack{\boldsymbol{c} \leq \boldsymbol{a} \\
\boldsymbol{d} \geq \boldsymbol{b} \\
D(\boldsymbol{c},\boldsymbol{d}) > 1}}
\frac{2\ln D(\boldsymbol{c},\boldsymbol{d})}
{d(\boldsymbol{c},\boldsymbol{d})},
\qquad d(\boldsymbol{c},\boldsymbol{d}) =
\max\limits_{i}f(i,\boldsymbol{c},\boldsymbol{d}) > 0,
\end{gathered}
\end{equation}
{\it then} $\beta(x,\boldsymbol{a}) \leq \beta(x,\boldsymbol{b})$.

P r o o f. Consider first the case $\boldsymbol{c}=\boldsymbol{a}$,
$\boldsymbol{d} = \boldsymbol{b}$. We have
\begin{equation}\label{prop11}
\begin{gathered}
\beta(x,\boldsymbol{a}) = \mathbf{P}
\left\{\boldsymbol{\xi} \in \mathcal A(x,\boldsymbol{a})\right\},
\qquad \mathcal A(x,\boldsymbol{a}) = \left\{{\mathbf y}:
\sum\limits_{i=1}^{n}a_{i}y_{i}^{2} \leq x \right\},
\end{gathered}
\end{equation}
where the ellipsoid
$\mathcal A(x,\boldsymbol{a}) \in \mathbb{R}^{n}$ has axes
$\{\sqrt{x/a_{i}},\ i=1,\ldots,n\}$ and volume
$V(\mathcal A(x,\boldsymbol{a}))$, proportional to
$\left(\prod\limits_{i=1}^{n}a_{i}\right)^{-1/2}$. In order to
compare probabilities $\beta(x,\boldsymbol{a})$ and
$\beta(x,\boldsymbol{b})$, consider the difference
(see \eqref{prop11})
$$
\begin{gathered}
(2\pi)^{n/2}\left[
\beta(x,\boldsymbol{a}) - \beta(x,\boldsymbol{b})\right] =
\Delta(x,\boldsymbol{a},\boldsymbol{b}) = \\
= \idotsint\limits_{\mathcal A(x,\boldsymbol{a})}
e^{-\frac{1}{2}\sum\limits_{i=1}^{n}y_{i}^{2}}d{\mathbf y} -
\idotsint\limits_{\mathcal A(x,\boldsymbol{b})}
e^{-\frac{1}{2}\sum\limits_{i=1}^{n}z_{i}^{2}}d{\mathbf z}.
\end{gathered}
$$
Changing variables
$z_{i} = \sqrt{a_{i}/b_{i}}\,y_{i}$, $i=1,\ldots,n$, in the second
integral we have ($D = D(\boldsymbol{a},\boldsymbol{b})$)
$$
\begin{gathered}
\Delta(x,\boldsymbol{a},\boldsymbol{b}) =
\idotsint\limits_{\mathcal A(x,\boldsymbol{a})}\left[
e^{-\frac{1}{2}\sum\limits_{i=1}^{n}y_{i}^{2}} -
D\exp\left\{-\frac{1}{2}\sum\limits_{j=1}^{n}\frac{a_{j}}{b_{j}}
y_{j}^{2}\right\}\right]d{\mathbf y} = \\
= \idotsint\limits_{\mathcal A(x,\boldsymbol{a})}
e^{-\frac{1}{2}\sum\limits_{i=1}^{n}y_{i}^{2}}
\left[1 - D\exp\left\{-\frac{1}{2}\sum\limits_{j=1}^{n}\left(
\frac{1}{b_{j}}- \frac{1}{a_{j}}\right)a_{j}
y_{j}^{2}\right\}\right]d{\mathbf y} \leq \\
\leq \left(1 - De^{-xd/2}\right)
\idotsint\limits_{\mathcal A(x,\boldsymbol{a})}
e^{-\frac{1}{2}\sum\limits_{i=1}^{n}y_{i}^{2}}d{\mathbf y}.
\end{gathered}
$$
Therefore $\Delta(x,\boldsymbol{a},\boldsymbol{b}) \leq 0$, if
$1 - De^{-xd/2} \leq 0$, i.e. if
$$
\begin{gathered}
x\max\limits_{i}f(i,\boldsymbol{a},\boldsymbol{b}) \leq
2\ln D = \ln\frac{W(\boldsymbol{a})}{W(\boldsymbol{b})} =
\sum\limits_{i=1}^{n}\ln\frac{a_{i}}{b_{i}},
\end{gathered}
$$
where $W(\boldsymbol{a}) = \prod\limits_{i=1}^{n}a_{i}$,
$W(\boldsymbol{b}) = \prod\limits_{i=1}^{n}b_{i}$, from which
Proposition 1 for $\boldsymbol{c}=\boldsymbol{a}$,
$\boldsymbol{d} = \boldsymbol{b}$ follows. If
$\boldsymbol{c} \leq \boldsymbol{a}$ and
$\boldsymbol{d} \geq \boldsymbol{b}$, then
$\beta(x,\boldsymbol{a}) \leq \beta(x,\boldsymbol{c})$ and
$\beta(x,\boldsymbol{d}) \leq \beta(x,\boldsymbol{b})$ for any
$x$ (see the statement 1 of Lemma below in section 3). Moreover,
$\beta(x,\boldsymbol{c}) \leq \beta(x,\boldsymbol{d})$, if
$x$ satisfies the condition \eqref{prop1a}. $\quad \blacktriangle$

{\it Remark} 2. Auxiliary vectors $\boldsymbol{c}$ and
$\boldsymbol{d}$ in \eqref{prop1a} allow sometimes to relax
the constraint \eqref{prop1a} in comparison with
$\boldsymbol{c} = \boldsymbol{a}$ and
$\boldsymbol{b} = \boldsymbol{d}$. One of further extensions
(Theorem 2) concerns the choice of vectors $\boldsymbol{c}$ and
$\boldsymbol{d}$ for given $\boldsymbol{a}$ and $\boldsymbol{b}$.

Below Proposition 1 is strengthened increasing the right-hand
side of the condition \eqref{prop1a} based on various
additional arguments.

{\bf 3. Strengthening 1}. We use the following auxiliary result.

L e m m a \ 1. 1) {\it Let
$\boldsymbol{a}, \boldsymbol{b} \in \mathbb{R}_{+}^{n}$ and for
components with indices $i,j$ we have
$\max\{a_{i},a_{j}\} \geq \max\{b_{i},b_{j}\}$
and $a_{i}a_{j} \geq b_{i}b_{j}$. Let also $a_{k} \geq b_{k}$
for all $k \neq i$, $k \neq j$. Then
$\beta(x,\boldsymbol{a}) \leq \beta(x,\boldsymbol{b})$
for any $x$.}

2) {\it If $\boldsymbol{b} \leq \boldsymbol{a}$, then
$\beta(x,\boldsymbol{a}) \leq \beta(x,\boldsymbol{b})$ for any
$x$.}

3) {\it Assume that it is possible to partition the set of indices
$I = \{1,2,\ldots,n\}$ on $k \geq 1$ parts  $I_{1},\ldots,I_{k}$,
such that
$I = \bigcup\limits_{j=1}^{k}I_{j}$, $I_{i}\cap I_{j}= \emptyset$,
$i \neq j$, and the following conditions are fulfilled
$$
\begin{gathered}
b_{i} \leq a_{0,j}, \qquad i \in I_{j}, \quad j=1,\ldots,k,
\end{gathered}
$$
where
$$
\begin{gathered}
a_{0,j}=
\left(\prod\limits_{i \in I_{j}}a_{i}\right)^{1/|I_{j}|}.
\end{gathered}
$$
Then $\beta(x,\boldsymbol{a}) \leq \beta(x,\boldsymbol{b})$
for any $x$.}

P r o o f. 1) Assume first that $n = 2$. Then the integration
region $\mathcal A(x,\boldsymbol{b}) \in \mathbb{R}^{2}$ for
$\beta(x,\boldsymbol{b})$ has volume
$V(\mathcal A(x,\boldsymbol{b}))$, proportional to
$(b_{1}b_{2})^{-1/2}$, and $V(\mathcal A(x,\boldsymbol{b})) \geq
V(\mathcal A(x,\boldsymbol{a}))$. The random vector
$\boldsymbol{\xi}=(\xi_{1},\xi_{2})$ has the distribution density
$p(\boldsymbol{y})$, proportional to
$e^{-r^{2}(\boldsymbol{y})/2}$,
$r^{2}(\boldsymbol{y}) = y_{1}^{2}+ y_{2}^{2}$,
monotonically decreasing in $r \geq 0$. It can be checked that
$$
\begin{gathered}
\inf_{\boldsymbol{y} \in \mathcal A(x,\boldsymbol{a}) \setminus
\mathcal A(x,\boldsymbol{b})} r^{2}(\boldsymbol{y}) \geq
\sup_{\boldsymbol{y} \in \mathcal A(x,\boldsymbol{b}) \setminus
\mathcal A(x,\boldsymbol{a})} r^{2}(\boldsymbol{y}).
\end{gathered}
$$
Therefore for given volume $V$ (i.e. for given product
$b_{1}b_{2} = T$) the value $\beta(x,\boldsymbol{b})$ attains its
maximum when $b_{1} = b_{2} = \sqrt{T}$, and monotonically decreases
when $b_{1}$ deviates from $\sqrt{T}$, from which the inequality
\eqref{Bak1} follows for any $x$.

If $n > 2$, then the inequality \eqref{Bak1} holds for any fixed
$\{\xi_{k}\}$, $k \neq i$, $k\neq j$ and any $x$, from which
necessary assertion follows.

2) If $\boldsymbol{b} \leq \boldsymbol{a}$, then
$\mathcal A(x,\boldsymbol{a}) \subseteq
\mathcal A(x,\boldsymbol{b})$, and therefore
$\beta(x,\boldsymbol{a}) \leq \beta(x,\boldsymbol{b})$ for any $x$.

3) That assertion follows from part 1). It is sufficient to
consider the case $k=2$, $I_{1} = \{1,\ldots,m\}$ and
$I_{2} = \{m+1,\ldots,n\}$ for some $1 < m < n$. Introduce
$n$-vector $\boldsymbol{a}_{0}=
(a_{0,1},\ldots,a_{0,1},a_{0,2},\ldots,a_{0,2})$, consisting of
$m$ components $a_{0,1}$ and $n-m$ components $a_{0,2}$. Then
repeatedly applying lemma's part 1) it is possible to show that
$\beta(x,\boldsymbol{a}) \leq \beta(x,\boldsymbol{a}_{0})$ for any
$x$. Since $\boldsymbol{b} \leq \boldsymbol{a}_{0}$, then
$\beta(x,\boldsymbol{a}_{0}) \leq \beta(x,\boldsymbol{b})$ for any
$x$. $\quad \blacktriangle$

We strengthen the Proposition 1. Setting for convenience
$\boldsymbol{c} = \boldsymbol{a}$ and
$\boldsymbol{d} = \boldsymbol{b}$, assume that there exists
$i \leq n-1$, such that $f(i+1,\boldsymbol{a},\boldsymbol{b}) <
f(i,\boldsymbol{a},\boldsymbol{b})$. Then $a_{i+1} < a_{i}$
(since $b_{i+1} \leq b_{i}$) and $b_{i+1} < b_{i}$. We build the
vector $\boldsymbol{a}^{(1)}$ such that
$\beta(x,\boldsymbol{a}) \leq \beta(x,\boldsymbol{a}^{(1)})$
for all $x$ and
$\beta(x,\boldsymbol{a}^{(1)}) \leq \beta(x,\boldsymbol{b})$
for $x$, satisfying, generally, a weaker than \eqref{prop1a}
constraint. For that purpose we use the following procedure.

Decrease $a_{i}$ down to the value $a_{i}^{(1)}$ and increase
respectively $a_{i+1}$ up to the value $a_{i+1}^{(1)}$, such that
the following three conditions are satisfied:

1) $a_{i}^{(1)} \geq a_{i+1}^{(1)}$;

2) $a_{i}a_{i+1}= a_{i}^{(1)}a_{i+1}^{(1)}$;

3) $f(i+1,\boldsymbol{a}^{(1)},\boldsymbol{b}) =
f(i,\boldsymbol{a}^{(1)},\boldsymbol{b})$, where
$\boldsymbol{a}^{(1)}$ - obtained that way from $\boldsymbol{a}$
the new monotone vector (it differs from
$\boldsymbol{a}$ only in components with indices $i$ and $i+1$).

Then we have $a_{i+1}^{(1)} < a_{i}^{(1)}$,
$f(i,\boldsymbol{a}^{(1)},\boldsymbol{b}) <
f(i,\boldsymbol{a},\boldsymbol{b})$ and
$f(i+1,\boldsymbol{a},\boldsymbol{b}) <
f(i+1,\boldsymbol{a}^{(1)},\boldsymbol{b})$. We also have
$\beta(x,\boldsymbol{a}) \leq \beta(x,\boldsymbol{a}^{(1)})$
for any $x$ (due to Lemma 1). Then after standard calculations
we have
$$
%\begin{equation}\label{stren14}
\begin{gathered}
f(i+1,\boldsymbol{a}^{(1)},\boldsymbol{b}) =
f(i,\boldsymbol{a}^{(1)},\boldsymbol{b}) = \frac{1}{2}\left[
a_{i}^{-1} + a_{i+1}^{-1}-\sqrt{
\left(a_{i}^{-1} + a_{i+1}^{-1}\right)^{2}}+z_{i}\right] <
f(i,\boldsymbol{a},\boldsymbol{b}),
\end{gathered}
$$
%\end{equation}
where
\begin{equation}\label{defuv1}
\begin{gathered}
z_{i}= 4\left(a_{i}^{-1}a_{i+1}^{-1}-b_{i}^{-1}b_{i+1}^{-1}\right),
\qquad i=1,\ldots,n.
\end{gathered}
\end{equation}
Note that values $f(i,\boldsymbol{a},\boldsymbol{b}),
f(i,\boldsymbol{a}^{(1)},\boldsymbol{b})$ may be negative.

Similarly, instead of the vector $\boldsymbol{a}$ we may change
the vector $\boldsymbol{b}$ (but in opposite direction), replacing
it by the vector $\boldsymbol{b}^{(1)}$ such that
$\beta(x,\boldsymbol{b}) \leq \beta(x,\boldsymbol{b}^{(1)})$
for all $x$ and
$\beta(x,\boldsymbol{a}) \leq \beta(x,\boldsymbol{b}^{(1)})$
for $x$, satisfying, generally, a weaker than \eqref{prop1a}
constraint. For that purpose, increase $b_{i}$ up to the value
$b_{i}^{(1)}$ and decrease respectively $b_{i+1}$ down to the value
$b_{i+1}^{(1)}$, such that the following three conditions
are satisfied:

1) $b_{i}^{(1)} \geq b_{i+1}^{(1)}$;

2) $b_{i}b_{i+1}= b_{i}^{(1)}b_{i+1}^{(1)}$;

3) $f(i+1,\boldsymbol{a},\boldsymbol{b}^{(1)}) =
f(i,\boldsymbol{a},\boldsymbol{b}^{(1)})$, where
$\boldsymbol{b}^{(1)}$ - obtained that way from $\boldsymbol{b}$
the new monotone vector (it differs from
$\boldsymbol{b}$ only in components with indices $i$ and $i+1$).

Then we have $b_{i+1}^{(1)} < b_{i}^{(1)}$,
$f(i,\boldsymbol{a},\boldsymbol{b}^{(1)}) <
f(i,\boldsymbol{a},\boldsymbol{b})$ and
$f(i+1,\boldsymbol{a},\boldsymbol{b}) <
f(i+1,\boldsymbol{a},\boldsymbol{b}^{(1)})$. We also have
$\beta(x,\boldsymbol{b}^{(1)}) \leq \beta(x,\boldsymbol{b})$
for any $x$ (due to Lemma). Then after standard calculations
we have
$$
\begin{gathered}
f(i+1,\boldsymbol{a},\boldsymbol{b}^{(1)}) =
f(i,\boldsymbol{a},\boldsymbol{b}^{(1)}) = \frac{1}{2}
\left(\sqrt{\left(a_{i}^{-1} + a_{i+1}^{-1}\right)^{2}- z_{i}}-
a_{i}^{-1} - a_{i+1}^{-1}\right) <
%u_{i,i+1}^{2}- z_{i}} - u_{i,i+1}\right) <
f(i,\boldsymbol{a},\boldsymbol{b}),
\end{gathered}
$$
where $z_{i}$ is defined in \eqref{defuv1}.

Compare values $f(i,\boldsymbol{a}^{(1)},\boldsymbol{b})$ and
$f(i,\boldsymbol{a},\boldsymbol{b}^{(1)})$. It is possible to check
that $f(i,\boldsymbol{a}^{(1)},\boldsymbol{b}) <
f(i,\boldsymbol{a},\boldsymbol{b}^{(1)})$ (i.e. changing the vector
$\boldsymbol{a}$ gives better result), if
$b_{i}b_{i+1} > a_{i}a_{i+1}$. If
$b_{i}b_{i+1} < a_{i}a_{i+1}$, then
$f(i,\boldsymbol{a}^{(1)},\boldsymbol{b}) >
f(i,\boldsymbol{a},\boldsymbol{b}^{(1)})$, i.e. changing the vector
$\boldsymbol{b}$ gives better result.

After that we apply the procedure described to obtained vectors
$\boldsymbol{a}^{(1)}$ or $\boldsymbol{b}^{(1)}$ and so on.
Unfortunately, the author was not able to investigate the optimal
sequence of changes the vectors $\boldsymbol{a}$ or
$\boldsymbol{b}$. For that reason we limit ourselves to the case
when only the vector $\boldsymbol{a}$ (or only the vector
$\boldsymbol{b}$) is changing.

Note that if $z_{i} \geq 0$ (i.e.
$b_{i}b_{i+1} \geq a_{i}a_{i+1}$), then $a_{i} \leq b_{i}$ and
$f(i,\boldsymbol{a},\boldsymbol{b}) \leq 0$, or
$a_{i+1} \leq b_{i+1}$ and
$f(i+1,\boldsymbol{a},\boldsymbol{b}) \leq 0$.

Denote by $i_{0}= i_{0}(\boldsymbol{a},\boldsymbol{b})$ the
minimal of indices $i$, such that
$f(i_{0},\boldsymbol{a},\boldsymbol{b}) =
\max\limits_{i}f(i,\boldsymbol{a},\boldsymbol{b})$, and by
$i_{1}= i_{1}(\boldsymbol{a},\boldsymbol{b}) \geq i_{0}$
the minimal of indices $i$, such that
$f(i_{1},\boldsymbol{a},\boldsymbol{b}) =
\max\limits_{i}f(i,\boldsymbol{a},\boldsymbol{b})$. If
$i_{1}(\boldsymbol{a},\boldsymbol{b}) =n$, then the method used
does not improve the condition \eqref{prop1a} and then in
Proposition 1 we have
$\max\limits_{i}f(i,\boldsymbol{a},\boldsymbol{b}) =
1/b_{n}-1/a_{n}$. Therefore we assume that
$i_{1}(\boldsymbol{a},\boldsymbol{b}) \leq n-1$.

{\it Change vector $\boldsymbol{a}$}.
First, for fixed monotone $\boldsymbol{b}$ we change the vector
$\boldsymbol{a}$. Choose arbitrary
$i \geq i_{0}(\boldsymbol{a},\boldsymbol{b})$, such that
$f(i+1,\boldsymbol{a},\boldsymbol{b}) <
f(i,\boldsymbol{a},\boldsymbol{b})$, and apply to $\boldsymbol{a}$
the procedure described. Then for obtained in that way from
$\boldsymbol{a}$ the new monotone vector $\boldsymbol{a}^{(1)}$
(it differs from $\boldsymbol{a}$ only in components with indices
$i$ and $i+1$) we have $a_{i+1}^{(1)} < a_{i}^{(1)}$,
$f(i,\boldsymbol{a}^{(1)},\boldsymbol{b}) <
f(i,\boldsymbol{a},\boldsymbol{b})$ and
$f(i+1,\boldsymbol{a},\boldsymbol{b}) <
f(i+1,\boldsymbol{a}^{(1)},\boldsymbol{b})$. We also have
$\beta(x,\boldsymbol{a}) \leq \beta(x,\boldsymbol{a}^{(1)})$
for any $x$ (due to Lemma).

As an initial index $i$ we may set, for example,
$i = i_{1}(\boldsymbol{a},\boldsymbol{b})$. Then apply the
procedure described to the received vector $\boldsymbol{a}^{(1)}$
(i.e. find a new index $i_{1}(\boldsymbol{a}^{(1)},\boldsymbol{b})$,
corresponding component $a_{i_{1}}^{(1)}$ and transform components
$a_{i_{1}}^{(1)}$ and $a_{i_{1}+1}^{(1)}$, such that three
conditions above are satisfied). It will give a new monotone vector
$\boldsymbol{a}^{(2)}$.  Then apply that procedure to the vector
$\boldsymbol{a}^{(2)}$ and so on.

As a result, we may get a sequence of monotone vectors
$\boldsymbol{a}^{m}$,  $m=1,2,\ldots$, converging to the
monotone vector $\boldsymbol{a}^{0}$. Let
$k_{1}$, $1 \leq k_{1} \leq i_{0}(\boldsymbol{a},\boldsymbol{b})$ -
minimal of indices $i$, which were used on all stages of getting
the sequence $\{\boldsymbol{a}^{m}\}$. Then for the limiting
monotone vector $\boldsymbol{a}^{0}$ the following conditions
are satisfied:
\begin{equation}\label{prop1am}
\begin{gathered}
f(i,\boldsymbol{a}^{0},\boldsymbol{b}) =
f(k_{1},\boldsymbol{a}^{0},\boldsymbol{b}),
\qquad k_{1} \leq i \leq n, \\
f(i,\boldsymbol{a}^{0},\boldsymbol{b}) <
f(k_{1},\boldsymbol{a}^{0},\boldsymbol{b}),
\qquad 1 \leq i \leq k_{1}-1, \\
\prod\limits_{i=k_{1}}^{n}a_{i}^{0} =
\prod\limits_{i=k_{1}}^{n}a_{i},
\end{gathered}
\end{equation}
i.e. the function $f(i,\boldsymbol{a}^{0},\boldsymbol{b})$ is
constant for $k_{1} \leq i \leq n$. Components
$\{a^{0}_{i}, i = 1,\ldots,k_{1}-1\}$ of the vector
$\boldsymbol{a}^{0}$ coincide with corresponding components
$\{a_{i}, i = 1,\ldots,k_{1}-1\}$ of the initial vector
$\boldsymbol{a}$ (they will not participate in getting the vector
$\boldsymbol{a}^{0}$).

We find the value $f(k_{1},\boldsymbol{a}^{0},\boldsymbol{b})$.
Since
$$
\begin{gathered}
a_{i}^{0} = \frac{b_{i}}
{1-f(k_{1},\boldsymbol{a}^{0},\boldsymbol{b})b_{i}},
\qquad k_{1} \leq i \leq n,
\end{gathered}
$$
then due to the last of conditions \eqref{prop1am} values
$\{f(i,\boldsymbol{a}^{0},\boldsymbol{b})\}$ satisfy also
equations
$$
\begin{gathered}
\prod\limits_{i=k_{1}}^{n}a_{i}^{0} = \prod\limits_{i=k_{1}}^{n}
\frac{b_{i}}{1-f(k_{1},\boldsymbol{a}^{0},\boldsymbol{b})b_{i}} =
\prod\limits_{i=k_{1}}^{n}a_{i},
\end{gathered}
$$
or, equivalently
\begin{equation}\label{prop1an}
\begin{gathered}
\sum\limits_{i=k_{1}}^{n}\ln\frac{a_{i}}{b_{i}} +
\sum\limits_{i=k_{1}}^{n}
\ln[1-f(k_{1},\boldsymbol{a}^{0},\boldsymbol{b})b_{i}] = 0.
\end{gathered}
\end{equation}

Define the value $T(k,\boldsymbol{a},\boldsymbol{b})$ as the
unique root of the equation
\begin{equation}\label{prop1am1}
\begin{gathered}
\sum\limits_{i=k}^{n}\ln\frac{a_{i}}{b_{i}} + \sum\limits_{i=k}^{n}
\ln[1-T(k,\boldsymbol{a},\boldsymbol{b})b_{i}] = 0.
\end{gathered}
\end{equation}
Then (see \eqref{prop1an} and \eqref{prop1am})
$T(k_{1},\boldsymbol{a},\boldsymbol{b}) =
T(k_{1},\boldsymbol{a}^{0},\boldsymbol{b}) =
f(k_{1},\boldsymbol{a}^{0},\boldsymbol{b})$. Note that if \\
$\sum\limits_{i=k_{1}}^{n}\ln(a_{i}/b_{i})> 0$,
then $T(k_{1},\boldsymbol{a},\boldsymbol{b}) > 0$, and if
$\sum\limits_{i=k_{1}}^{n}\ln(a_{i}/b_{i})< 0$, then
$T(k_{1},\boldsymbol{a},\boldsymbol{b})< 0$.

The index $k_{1}(\boldsymbol{a},\boldsymbol{b})$,
$1 \leq k_{1} \leq i_{0}(\boldsymbol{a},\boldsymbol{b})$ is
defined as follows
\begin{equation}\label{defk1}
\begin{gathered}
k_{1}(\boldsymbol{a},\boldsymbol{b}) =
\min\left\{k:f(k,\boldsymbol{a},\boldsymbol{b}) \geq
T(k,\boldsymbol{a},\boldsymbol{b})\right\} = \\
= \max\left\{k:f(k-1,\boldsymbol{a},\boldsymbol{b}) <
T(k,\boldsymbol{a},\boldsymbol{b})\right\}.
\end{gathered}
\end{equation}

Consider now first $k_{1}-1$ coordinates $a^{0}_{i}$,
$i=1,\ldots,k_{1}-1$ of the vector $\boldsymbol{a}^{0}$. Again
we find for them corresponding values $i_{0}', i_{1}',k_{1}'$
and replace coordinates $a^{0}_{i}$, $i=k_{1}',\ldots,k_{1}-1$ by
corresponding coordinates $a^{0'}_{i}$, $i=k_{1}',\ldots,k_{1}-1$,
such that the function $f(i,\boldsymbol{a}^{0},\boldsymbol{b})$
is constant for $k_{1}' \leq i \leq k_{1}-1$. Continuing that
process, we get the monotone vector $\boldsymbol{a}^{(0)}$,
such that
$\beta(x,\boldsymbol{a}) \leq \beta(x,\boldsymbol{a}^{(0)})$
for any $x$. Moreover, the function
$f(i,\boldsymbol{a}^{(0)},\boldsymbol{b})$ is piecewise constant
and does not decrease in $i$.

{\it Change vector $\boldsymbol{b}$}. Similarly for fixed
$\boldsymbol{a}$ we may sequentially change the vector
$\boldsymbol{b}$ (but in opposite direction), again using the index
$i_{1}(\boldsymbol{a},\boldsymbol{b})$ and getting the sequence of
monotone vectors $\{\boldsymbol{b}^{(i)}\}$, $i=1,2,\ldots$,
such that
$\beta(x,\boldsymbol{b}) \geq \beta(x,\boldsymbol{b}^{(1)}) \geq
\beta(x,\boldsymbol{b}^{(2)}) \geq \ldots$ for any $x$ (because of
Lemma). As in the case of vector $\boldsymbol{a}$, assume that
$i_{1}(\boldsymbol{a},\boldsymbol{b}) \leq n-1$. Then
$f(i_{1}+1,\boldsymbol{a},\boldsymbol{b}) <
f(i_{1},\boldsymbol{a},\boldsymbol{b})$ and
$b_{i_{1}+1} < b_{i_{1}}$. Apply to $\boldsymbol{b}$ the procedure
described and get new monotone vector $\boldsymbol{b}^{(1)}$
(it differs from $\boldsymbol{b}$ only in components with indices
$i$ and $i+1$). We have for it
$b_{i_{1}+1}^{(1)} < b_{i_{1}}^{(1)}$,
$f(i_{1},\boldsymbol{a},\boldsymbol{b}^{(1)}) <
f(i_{1},\boldsymbol{a},\boldsymbol{b})$ and
$f(i_{1}+1,\boldsymbol{a},\boldsymbol{b}) <
f(i_{1}+1,\boldsymbol{a},\boldsymbol{b}^{(1)})$. We also have
$\beta(x,\boldsymbol{b}^{(1)}) \leq \beta(x,\boldsymbol{b})$
for any $x$ (because of Lemma).

Then apply the procedure described to the obtained vector
$\boldsymbol{b}^{(1)}$ (i.e. find the new index
$i_{1}(\boldsymbol{a},\boldsymbol{b}^{(1)})$ and corresponding
component $b_{i_{1}}^{(1)}$ and transform components
$b_{i_{1}}^{(1)}$ and $b_{i_{1}+1}^{(1)}$, such that three
conditions above are satisfied). It will give the vector
$\boldsymbol{b}^{(2)}$. Then apply that procedure to the vector
$\boldsymbol{b}^{(2)}$ and so on.

As a result, we may get a sequence of monotone vectors
$\boldsymbol{b}^{m}$,  $m=1,2,\ldots$, converging to the
monotone vector $\boldsymbol{b}^{0}$. Let
$k_{2}$, $1 \leq k_{2} \leq i_{1}(\boldsymbol{a},\boldsymbol{b})$ -
minimal of indices $i_{1}$, used on all stages of getting
the sequence $\{\boldsymbol{b}^{m}\}$. Then for the limiting
monotone vector $\boldsymbol{b}^{0}$ the following conditions,
similar to \eqref{prop1am}, are satisfied:
\begin{equation}\label{prop1ab}
\begin{gathered}
f(i,\boldsymbol{a},\boldsymbol{b}^{(0)}) =
f(k_{2},\boldsymbol{a},\boldsymbol{b}^{(0)}),
\qquad k_{2} \leq i \leq n, \\
f(i,\boldsymbol{a},\boldsymbol{b}^{(0)}) <
f(k_{2},\boldsymbol{a},\boldsymbol{b}^{(0)}),
\qquad 1 \leq i \leq k_{2}-1, \\
\prod\limits_{i=1}^{n}b_{i}^{(0)} = \prod\limits_{i=1}^{n}b_{i},
\end{gathered}
\end{equation}
i.e. the function $f(i,\boldsymbol{a},\boldsymbol{b}^{(0)})$ is
constant for $k_{2} \leq i \leq n$. Components
$\{b^{(0)}_{i}, i = 1,\ldots,k_{2}-1\}$ of the vector
$\boldsymbol{b}^{(0)}$ coincide with corresponding components
$\{b_{i}, i = 1,\ldots,k_{2}-1\}$ of the initial vector
$\boldsymbol{b}$ (they will not participate in getting the vector
$\boldsymbol{b}^{0}$). Now if
$\beta(x,\boldsymbol{a}) \leq \beta(x,\boldsymbol{b}^{(0)})$,
then $\beta(x,\boldsymbol{a}) \leq \beta(x,\boldsymbol{b})$.

Since
$$
\begin{gathered}
b_{i}^{(0)} = \frac{a_{i}}
{1+f(i,\boldsymbol{a},\boldsymbol{b}^{(0)})a_{i}},
\qquad k_{2} \leq i \leq n,
\end{gathered}
$$
then due to the last of conditions \eqref{prop1ab} values
$\{f(i,\boldsymbol{a},\boldsymbol{b}^{(0)})\}$ satisfy also
equations
$$
\begin{gathered}
\prod\limits_{i=k_{2}}^{n}b_{i}^{(0)} = \prod\limits_{i=k_{2}}^{n}
\frac{a_{i}}{1+f(k_{2},\boldsymbol{a},\boldsymbol{b}^{(0)})a_{i}} =
\prod\limits_{i=k_{2}}^{n}b_{i},
\end{gathered}
$$
or, equivalently
$$
\begin{gathered}
\sum\limits_{i=k_{2}}^{n}\ln\frac{a_{i}}{b_{i}} -
\sum\limits_{i=k_{2}}^{n}
\ln[1+f(k_{2},\boldsymbol{a},\boldsymbol{b}^{(0)})a_{i}] = 0.
\end{gathered}
$$

Define the value $D(k,\boldsymbol{a},\boldsymbol{b})$ as the
unique root of the equation
\begin{equation}\label{prop1am2}
\begin{gathered}
\sum\limits_{i=k}^{n}\ln\frac{a_{i}}{b_{i}} - \sum\limits_{i=k}^{n}
\ln[1+D(k,\boldsymbol{a},\boldsymbol{b})a_{i}] = 0.
\end{gathered}
\end{equation}
Then (see \eqref{prop1an} and \eqref{prop1am})
$D(k_{2},\boldsymbol{a},\boldsymbol{b}) =
D(k_{2},\boldsymbol{a},\boldsymbol{b}^{(0)}) =
f(k_{2},\boldsymbol{a},\boldsymbol{b}^{(0)})$.
Note that if \\
$\sum\limits_{i=k_{2}}^{n}\ln(a_{i}/b_{i})> 0$,
then $D(k_{2},\boldsymbol{a},\boldsymbol{b}) > 0$, and if
$\sum\limits_{i=k_{2}}^{n}\ln(a_{i}/b_{i})< 0$, then
$D(k_{2},\boldsymbol{a},\boldsymbol{b})< 0$.

The index $k_{2}(\boldsymbol{a},\boldsymbol{b})$,
$1 \leq k_{2} \leq i_{0}(\boldsymbol{a},\boldsymbol{b})$ can be
defined as follows
\begin{equation}\label{defk2}
\begin{gathered}
k_{2}(\boldsymbol{a},\boldsymbol{b}) =
\min\left\{k:f(k,\boldsymbol{a},\boldsymbol{b}) \geq
D(k,\boldsymbol{a},\boldsymbol{b})\right\} = \\
= \max\left\{k:f(k-1,\boldsymbol{a},\boldsymbol{b}) <
D(k,\boldsymbol{a},\boldsymbol{b})\right\}.
\end{gathered}
\end{equation}

In order to formulate the result obtained note that similarly
to Proposition 1 we may additionally introduce arbitrary vectors
$\boldsymbol{c} \leq \boldsymbol{a}$ and
$\boldsymbol{d} \geq \boldsymbol{b}$, such that
$D(\boldsymbol{c},\boldsymbol{d}) > 1$. Then we have

T h e o r e m \,1. {\it Let
$\boldsymbol{a}, \boldsymbol{b} \in \mathbb{R}^{n}_{+}$ - monotone
decreasing vectors. If $x$ satisfies condition}
\begin{equation}\label{prop1ad}
\begin{gathered}
x \leq
\sup_{\substack{\boldsymbol{c} \leq \boldsymbol{a} \\
\boldsymbol{d} \geq \boldsymbol{b} \\
D(\boldsymbol{c},\boldsymbol{d}) > 1}}
\left\{
\frac{2\ln D(\boldsymbol{c},\boldsymbol{d})}{
\min\left\{T(k_{1},\boldsymbol{c},\boldsymbol{d}),
D(k_{2},\boldsymbol{c},\boldsymbol{d})\right\}}
\right\},
\end{gathered}
\end{equation}
({\it values $T(k,\boldsymbol{c},\boldsymbol{d}) and
D(k,\boldsymbol{c},\boldsymbol{d})$ are defined in
\eqref{prop1am1} and \eqref{prop1am2}, and values
$k_{1}=k_{1}(\boldsymbol{c},\boldsymbol{d})$ and
$k_{2}=k_{2}(\boldsymbol{c},\boldsymbol{d})$  in \eqref{defk1} and
\eqref{defk2}})
{\it then} $\beta(x,\boldsymbol{a}) \leq \beta(x,\boldsymbol{b})$.

{\it Remark} 3. If
$$
\begin{gathered}
\frac{1}{b_{1}}- \frac{1}{a_{1}} =
\max\limits_{i}\left(\frac{1}{b_{i}}- \frac{1}{a_{i}}\right),
\end{gathered}
$$
then $k_{1}(\boldsymbol{a},\boldsymbol{b}) =
k_{2}(\boldsymbol{a},\boldsymbol{b}) = 1$ and
$$
\begin{gathered}
T(k_{1},\boldsymbol{a},\boldsymbol{b}) =
T(1,\boldsymbol{a},\boldsymbol{b}), \qquad
D(k_{2},\boldsymbol{a},\boldsymbol{b}) =
D(1,\boldsymbol{a},\boldsymbol{b}).
\end{gathered}
$$

The value $T(k)=T(k,\boldsymbol{a},\boldsymbol{b})$ can be
upper bounded, for example, as follows. Using Jensen's inequality
$\mathbf{E}\ln \xi \leq \ln \mathbf{E}\xi$,
from \eqref{prop1am1} we have
$$
\begin{gathered}
\sum\limits_{i=k}^{n}\ln\frac{a_{i}}{b_{i}} =-\sum\limits_{i=k}^{n}
\ln[1-T(k)b_{i}] \geq -(n-k+1)\ln\left[1-\frac{T(k)
\sum\limits_{i=k}^{n}b_{i}}{(n-k+1)}\right].
\end{gathered}
$$
Applying the inequality $\ln(1+x) \leq x$, we get
\begin{equation}\label{prop1ba}
T(k,\boldsymbol{a},\boldsymbol{b}) \leq
\frac{(n-k+1)\left[1-\left(\prod\limits_{i=k}^{n}
\dfrac{b_{i}}{a_{i}}\right)^{1/(n-k+1)}\right]}
{\sum\limits_{i=k}^{n}b_{i}} \leq
\frac{\sum\limits_{i=k}^{n}\ln\dfrac{a_{i}}{b_{i}}}
{\sum\limits_{i=k}^{n}b_{i}}.
\end{equation}
For the value $D(k,\boldsymbol{a},\boldsymbol{b})$ those estimates
work in opposite direction:
$$
D(k,\boldsymbol{a},\boldsymbol{b}) \geq
\frac{\sum\limits_{i=k}^{n}\ln\dfrac{a_{i}}{b_{i}}}
{\sum\limits_{i=k}^{n}a_{i}}.
$$

C o r o l l a r y \,1. {\it If $a_{1} \geq b_{1}$,
the inequality holds
\begin{equation}\label{cor1}
\begin{gathered}
\sum\limits_{i=1}^{n}\ln\frac{a_{i}}{b_{i}} \geq 0
\end{gathered}
\end{equation}
and any of the following conditions is valid
\begin{equation}\label{cor1a}
\begin{gathered}
\sum\limits_{i=1}^{n}\ln\frac{a_{i}}{b_{i}} +
\sum\limits_{i=1}^{n}
\ln\left[1-\left(\frac{1}{b_{1}}- \frac{1}{a_{1}}\right)b_{i}\right]
\leq 0,
\end{gathered}
\end{equation}
\begin{equation}\label{cor1b}
\begin{gathered}
\sum\limits_{i=1}^{n}\ln\frac{a_{i}}{b_{i}} -
\sum\limits_{i=1}^{n}
\ln\left[1+\left(\frac{1}{b_{1}}- \frac{1}{a_{1}}\right)a_{i}\right]
\leq 0,
\end{gathered}
\end{equation}
then
$\beta(x,\boldsymbol{a}) \leq \beta(x,\boldsymbol{b})$ for
any $x$}.

P r o o f. We set $\boldsymbol{c} = \boldsymbol{a}$ and
$\boldsymbol{b} = \boldsymbol{d}$. Show that if $a_{1} \geq b_{1}$,
and conditions \eqref{cor1} and \eqref{cor1a} are satisfied,
then $k_{1}(\boldsymbol{a},\boldsymbol{b}) = 1$. Indeed, due to
\eqref{defk1} for that purpose it is sufficient to have
$$
\begin{gathered}
f(1,\boldsymbol{a},\boldsymbol{b}) = \frac{1}{b_{1}} -
\frac{1}{a_{1}} \geq T(1,\boldsymbol{a},\boldsymbol{b}),
\end{gathered}
$$
where $T = T(1,\boldsymbol{a},\boldsymbol{b}) \geq 0$
(see \eqref{prop1am1}) -- the unique root of the equation
\begin{equation}\label{prop1am11}
\begin{gathered}
F(T) = \sum\limits_{i=1}^{n}\ln\frac{a_{i}}{b_{i}} +
\sum\limits_{i=1}^{n}\ln(1-Tb_{i}) = 0.
\end{gathered}
\end{equation}
Since $F(0) \geq 0$ and $F'(T) < 0$, then the equation
\eqref{prop1am11} has the unique root
$T=T(1,\boldsymbol{a},\boldsymbol{b}) \geq 0$. Moreover,
if $F[f(1,\boldsymbol{a},\boldsymbol{b})] \leq 0$, then
$f(1,\boldsymbol{a},\boldsymbol{b}) \geq
T(1,\boldsymbol{a},\boldsymbol{b})$ (and then
$k_{1}(\boldsymbol{a},\boldsymbol{b}) = 1$), which coincides with
the condition \eqref{cor1a}. We also have
\begin{equation}\label{cor1am}
\begin{gathered}
f(i,\boldsymbol{a}^{(0)},\boldsymbol{b}) = \frac{1}{b_{i}} -
\frac{1}{a_{i}^{(0)}} = \frac{1}{b_{1}} -\frac{1}{a_{1}^{(0)}} =
T(1,\boldsymbol{a},\boldsymbol{b}) \geq 0,
\qquad 1 \leq i \leq n.
\end{gathered}
\end{equation}
Hence $\boldsymbol{b} \leq \boldsymbol{a}^{(0)}$, and therefore
$\beta(x,\boldsymbol{a}) \leq \beta(x,\boldsymbol{a}^{(0)})
\leq \beta(x,\boldsymbol{b})$ for any $x$.

Similarly we can show that if $a_{1} \geq b_{1}$, and also
conditions \eqref{cor1} and \eqref{cor1b} are satisfied, then
$k_{2}(\boldsymbol{a},\boldsymbol{b}) = 1$, and therefore
$\beta(x,\boldsymbol{a}) \leq \beta(x,\boldsymbol{b}^{(0)})
\leq \beta(x,\boldsymbol{b})$ for any $x$. $\quad \blacktriangle$

The following result is also valid.

P r o p o s i t i o n \,2. {\it Let
$\boldsymbol{a}, \boldsymbol{b} \in \mathbb{R}^{n}_{+}$. If
$\beta(x,\boldsymbol{a}) \leq \beta(x,\boldsymbol{b})$
for any $x > 0$, then
$\max\limits_{i}a_{i} \geq \max\limits_{i}b_{i}$.}

P r o o f. It is sufficient to consider the case
$\max\limits_{i}a_{i} = a_{1}$, $\max\limits_{i}b_{i} = b_{1}$,
where $b_{1} > 0$. Assume that $a_{1} < b_{1}$. Show that then the
inequality $\beta(x,\boldsymbol{a}) \leq \beta(x,\boldsymbol{b})$
can not hold for large $x$. Denoting
$\boldsymbol{a}_{1} = (a_{1},\ldots,a_{1})$ and
$\boldsymbol{b}_{1} = (b_{1},0,\ldots,0)$, note that
$\beta(x,\boldsymbol{a}_{1}) \leq \beta(x,\boldsymbol{a})$ and
$\beta(x,\boldsymbol{b}) \leq \beta(x,\boldsymbol{b}_{1})$.
Therefore it is sufficient to show that for large $x$
the inequality
$\beta(x,\boldsymbol{a}_{1}) \leq \beta(x,\boldsymbol{b}_{1})$
does not hold. We may limit ourselves to the case
$b_{1} = 1$, $a_{1} < 1$. Then the inequality,
equivalent to \eqref{Bak1}, takes the form
\begin{equation}\label{prop2}
1-\beta(x,\boldsymbol{a}_{1}) = \mathbf{P}
\left(a_{1}\sum\limits_{i=1}^{n}\xi_{i}^{2}> x\right) \geq
\mathbf{P}\left(\xi_{1}^{2}> x\right) = 1-
\beta(x,\boldsymbol{b}_{1}).
\end{equation}
The left side of that formula can be bounded using exponential
Chebychev inequality
$$
\mathbf{P}\left(a_{1}\sum\limits_{i=1}^{n}\xi_{i}^{2}< x\right)
\leq \exp\left\{-\frac{1}{2}\left(n\ln\frac{na_{1}}{ex} +
\frac{x}{a_{1}}\right)\right\}, \qquad x \geq a_{1}n,
$$
and therefore
$$
\ln\mathbf{P}\left(a_{1}\sum\limits_{i=1}^{n}\xi_{i}^{2}< x\right)
\leq -\frac{x+ o(x)}{2a_{1}}, \qquad x \to \infty.
$$

On the other hand, using the standard estimate
$$
\mathbf{P}\left(\xi_{1}^{2}> x\right) =
2\mathbf{P}\left(\xi_{1} > \sqrt{x}\right) \geq
\frac{2\sqrt{x}}{\sqrt{2\pi}(1+x)}e^{-x/2}, \qquad x \geq 0,
$$
we have
$$
\begin{gathered}
\ln\mathbf{P}\left(\xi_{1}^{2}> x\right) \geq
-\frac{x+ o(x)}{2} >
-\frac{x+ o(x)}{2a_{1}} \geq
\ln\mathbf{P}\left(a_{1}\sum\limits_{i=1}^{n}\xi_{i}^{2}< x\right),
\qquad x \to \infty,
\end{gathered}
$$
from which it follows that the inequality \eqref{prop2} can not
hold for large $x$. $\quad \blacktriangle$

From remark 3 and the estimate \eqref{prop1ba} we also get

C o r o l l a r y \,2. {\it Let
$\boldsymbol{a}, \boldsymbol{b} \in \mathbb{R}_{+}^{n}$ -
monotone vectors, such that
$$
\begin{gathered}
\frac{1}{b_{1}}- \frac{1}{a_{1}} =
\max\limits_{i}\left(\frac{1}{b_{i}}- \frac{1}{a_{i}}\right), \\
\prod\limits_{i=1}^{n}a_{i} \geq \prod\limits_{i=1}^{n}b_{i}.
\end{gathered}
$$
Then }
$$
%\begin{eqnarray}\label{Cor1}
\beta(x,\boldsymbol{a}) \leq \beta(x,\boldsymbol{b}), \qquad
x \leq \sum\limits_{i=1}^{n}b_{i}.
$$
%\end{eqnarray}

{\bf 4. Strengthening 2}. For vectors
$\boldsymbol{a}, \boldsymbol{b} \in \mathbb{R}_{+}^{n}$ we use
the function $f(i,\boldsymbol{a},\boldsymbol{b})$ from
\eqref{deff1}. In each of vectors
$\boldsymbol{a}, \boldsymbol{b}$ we change in some way
numeration of their components $\{i\}$, such that the function
$f(i,\boldsymbol{a},\boldsymbol{b})$ become monotone increasing in
$i$ and, in particular,
$$
\begin{gathered}
\max_{i}f(i,\boldsymbol{a},\boldsymbol{b})=
b_{n}^{-1} - a_{n}^{-1}.
\end{gathered}
$$
Of course, then the values
$\beta(x,\boldsymbol{a}), \beta(x,\boldsymbol{b})$ will not be
changed. It is desirable to have the value
$b_{n}^{-1} - a_{n}^{-1}$ as small as possible.

Then for $\boldsymbol{c}=\boldsymbol{a}$,
$\boldsymbol{d} = \boldsymbol{b}$ the condition \eqref{prop1a}
takes the form
\begin{equation}\label{stren1a}
\begin{gathered}
x \leq G(n,\boldsymbol{a},\boldsymbol{b}) =
\frac{1}{d}\sum\limits_{i=1}^{n}\ln\frac{a_{i}}{b_{i}},
\qquad d = b_{n}^{-1} - a_{n}^{-1}.
\end{gathered}
\end{equation}

Note that if in \eqref{stren1a}
$\sum\limits_{i=1}^{n}\ln(a_{i}/b_{i}) > 0$, then $d > 0$. Denote
\begin{equation}\label{stren1b}
\begin{gathered}
t_{1} = t_{1}(\boldsymbol{a},\boldsymbol{b}) =
\min\{i: b_{i}^{-1} - a_{i}^{-1} = b_{n}^{-1} - a_{n}^{-1}\}.
\end{gathered}
\end{equation}
Then $f(i,\boldsymbol{a},\boldsymbol{b}) =
f(t_{1},\boldsymbol{a},\boldsymbol{b}) =
b_{t_{1}}^{-1} - a_{t_{1}}^{-1}$, $t_{1} \leq i \leq n$.

We try to increase the right-hand side of \eqref {stren1a},
increasing $b_{i}$, $t_{1}\leq i \leq n$, but not changing
$\{a_{i}\}$. Then the value $d$ will decrease. For some
$\varepsilon \geq 0$ we set
$$
\begin{gathered}
(b_{i}')^{-1} = b_{i}^{-1} - \varepsilon, \quad i=t_{1},\ldots,n;
\qquad  b_{i}' = b_{i}, \quad i < t_{1}, \qquad
\boldsymbol{b}' = (b_{1}',\ldots,b_{n}').
\end{gathered}
$$
Then
$$
\begin{gathered}
G(n,\boldsymbol{a},\boldsymbol{b}') =
\frac{1}{(d - \varepsilon)}\left[\sum\limits_{i=1}^{t_{1}-1}
\ln\frac{a_{i}}{b_{i}} + \sum\limits_{i=t_{1}}^{n}
\ln\left[a_{i}(b_{i}^{-1}-\varepsilon)\right]\right]
\end{gathered}
$$
and (since $\ln z \geq 1-1/z$)
$$
\begin{gathered}
G'_{\varepsilon}(n,\boldsymbol{a},\boldsymbol{b}') =
\frac{1}{(d - \varepsilon)^{2}}\left[\sum\limits_{i=1}^{t_{1}-1}
\ln\frac{a_{i}}{b_{i}} + \sum\limits_{i=t_{1}}^{n}
\ln\left[a_{i}(b_{i}^{-1}-\varepsilon)\right]\right] -
\frac{1}{(d - \varepsilon)}\sum\limits_{i=t_{1}}^{n}
\frac{1}{b_{i}^{-1}-\varepsilon} = \\
= \frac{1}{(d - \varepsilon)^{2}}\left[\sum\limits_{i=1}^{t_{1}-1}
\ln\frac{a_{i}}{b_{i}} + \sum\limits_{i=t_{1}}^{n}\left(
\ln\left[a_{i}(b_{i}^{-1}-\varepsilon)\right] - 1 +
\frac{1}{a_{i}(b_{i}^{-1}-\varepsilon)}\right)\right] \geq \\
\geq \frac{1}{(d - \varepsilon)^{2}}\sum\limits_{i=1}^{t_{1}-1}
\ln\frac{a_{i}}{b_{i}}.
\end{gathered}
$$

Therefore if
$\sum\limits_{i=1}^{t_{1}-1}\ln\dfrac{a_{i}}{b_{i}} > 0$, then
decrease all $b_{i}^{-1}$, $t_{1} \leq i \leq n$, on the value
$\varepsilon$, such that for the new vector
$\boldsymbol{b}' \geq \boldsymbol{b}$ (and then
$\beta(x,\boldsymbol{b}') \leq \beta(x,\boldsymbol{b})$ for any
$x$) we will have
$$
f(i,\boldsymbol{a},\boldsymbol{b}') = (b_{i}')^{-1} - a_{i}^{-1} =
f(t_{1}-1,\boldsymbol{a},\boldsymbol{b}) =
b_{t_{1}-1}^{-1}-a_{t_{1}-1}^{-1}, \qquad t_{1}-1 \leq i \leq n,
$$
i.e. the function $f(i,\boldsymbol{a},\boldsymbol{b}')$ becomes
constant for $t_{1}-1 \leq i \leq n$. For that purpose we set
$$
\varepsilon= b_{n}^{-1}-a_{n}^{-1} -
(b_{t_{1}-1}^{-1}-a_{t_{1}-1}^{-1}) > 0
$$
and then get
$$
\begin{gathered}
G(n,\boldsymbol{a},\boldsymbol{b}') = \frac{1}{d'}
\sum\limits_{i=1}^{n}\ln\frac{a_{i}}{b_{i}'} >
G(n,\boldsymbol{a},\boldsymbol{b}), \qquad
d' = b_{t_{1}-1}^{-1}-a_{t_{1}-1}^{-1} > 0, \\
(b_{i}')^{-1}= a_{i}^{-1} + b_{t_{1}-1}^{-1}-a_{t_{1}-1}^{-1},
\quad i=t_{1}-1,\ldots,n; \qquad b_{i}' = b_{i}, \quad
i=1,\ldots,t_{1}-2.
\end{gathered}
$$

Similarly we repeat that procedure for the obtained vector
$\boldsymbol{b}'$. Denote
$$
\begin{gathered}
t_{2} = t_{2}(\boldsymbol{a},\boldsymbol{b}) =
t_{1}(\boldsymbol{a},\boldsymbol{b}') =
\min\{i: b_{i}^{-1}-a_{i}^{-1} =
b_{t_{1}-1}^{-1}-a_{t_{1}-1}^{-1}\}.
\end{gathered}
$$
Next, again if
$\sum\limits_{i=1}^{t_{2}-1}\ln\dfrac{a_{i}}{b_{i}} > 0$, then
decrease all $(b_{i}')^{-1}$, $t_{2} \leq i \leq n$ on the value
$\varepsilon_{1}$, such that for the new vector
$\boldsymbol{b}'' \geq \boldsymbol{b}'$ we have
$$
f(i,\boldsymbol{a},\boldsymbol{b}'') =(b_{i}'')^{-1} - a_{i}^{-1}=
f(t_{2}-1,\boldsymbol{a},\boldsymbol{b})=
b_{t_{2}-1}^{-1}-a_{t_{2}-1}^{-1}, \qquad
t_{2}-1 \leq i \leq n,
$$
i.e. the function $f(i,\boldsymbol{a},\boldsymbol{b}'')$ become
constant for $t_{2}-1 \leq i \leq n$. For that purpose we set
$$
\varepsilon_{1} = b_{n}^{-1}-a_{n}^{-1} -
(b_{t_{2}-1}^{-1}-a_{t_{2}-1}^{-1}) > 0.
$$
Then we get
$$
\begin{gathered}
G(n,\boldsymbol{a},\boldsymbol{b}'') = \frac{1}{d''}
\sum\limits_{i=1}^{n}\ln\frac{a_{i}}{b_{i}''} >
G(n,\boldsymbol{a},\boldsymbol{b}'), \qquad
d'' = b_{t_{2}-1}^{-1}-a_{t_{2}-1}^{-1} > 0, \\
(b_{i}'')^{-1}= a_{i}^{-1} + b_{t_{2}-1}^{-1}-a_{t_{2}-1}^{-1},
\quad i=t_{2}-1,\ldots,n; \qquad b_{i}'' = b_{i}, \quad
i=1,\ldots,t_{2}-2.
\end{gathered}
$$
Repeating that process, we get the following result. Denote
\begin{equation}\label{stren1e}
\begin{gathered}
n_{1} = n_{1}(\boldsymbol{a},\boldsymbol{b}) = \min\left\{m:
\sum\limits_{i=1}^{j}\ln\dfrac{a_{i}}{b_{i}} \geq 0, \quad
j=m,\ldots,n \right\}.
\end{gathered}
\end{equation}
Then the following result holds (see \eqref{stren1a}).

T h e o r e m \,2. {\it Let
$\boldsymbol{a}, \boldsymbol{b} \in \mathbb{R}_{+}^{n}$ and let
the function $f(i,\boldsymbol{a},\boldsymbol{b})$ monotonically
increases in $i$.}

1) {\it If $n_{1}=n_{1}(\boldsymbol{a},\boldsymbol{b}) \geq 2$
and \ $x$ satisfies the condition
\begin{equation}\label{stren22}
\begin{gathered}
x \leq G(n,\boldsymbol{a},\boldsymbol{b}),
\end{gathered}
\end{equation}
where}
\begin{equation}\label{stren2}
\begin{gathered}
G(n,\boldsymbol{a},\boldsymbol{b}) = \frac{1}{d}
\left(\sum\limits_{i=1}^{n_{1}-1}\ln\frac{a_{i}}{b_{i}} +
\sum\limits_{i=n_{1}}^{n}\ln(1+da_{i})\right),
\qquad d = b_{n_{1}}^{-1}-a_{n_{1}}^{-1},
\end{gathered}
\end{equation}
{\it then} $\beta(x,\boldsymbol{a}) \leq \beta(x,\boldsymbol{b})$.

2) {\it If $n_{1}(\boldsymbol{a},\boldsymbol{b}) = 1$, then
$\beta(x,\boldsymbol{a}) \leq \beta(x,\boldsymbol{b})$ for all}
$x$.

Explain only the statement 2 of that Theorem. If
$n_{1}(\boldsymbol{a},\boldsymbol{b}) = 1$, then $a_{1}\geq b_{1}$.
Since the function $f(i,\boldsymbol{a},\boldsymbol{b})$
monotonically increases in $i$, then $a_{i} \geq b_{i}$ for all
$i$, and therefore $\boldsymbol{a} \geq \boldsymbol{b}$. Then from
the statement 2 of Lemma we get
$\beta(x,\boldsymbol{a}) \leq \beta(x,\boldsymbol{b})$ for all $x$.

{\bf 5. Examples}. 1. Let $\boldsymbol{b} = \boldsymbol{1}$. If
$\prod\limits_{i=1}^{n}a_{i} \geq 1$, then for any $x$ we have
(see the statement 3 of Lemma)
$\beta(x,\boldsymbol{a}) \leq \beta(x,\boldsymbol{1})$, i.e.
\begin{eqnarray}\label{Cor21}
\mathbf{P}\left(\sum\limits_{i=1}^{n}a_{i}\xi_{i}^{2}< x\right)
\leq \mathbf{P}\left(\sum\limits_{i=1}^{n}\xi_{i}^{2}< x\right).
\end{eqnarray}

Compare results which give in that example \cite[Theorem 1]{Bak95},
Proposition 1 and Theorems 1, 2 of the paper.

If additionally the condition \eqref{Bak2} is also satisfied then
according to \cite[Theorem 1]{Bak95} the inequality \eqref{Cor21}
holds for $x \geq 2n$.

According to Proposition 1 (i.e. the condition \eqref{prop1a}) for
$\boldsymbol{c} = \boldsymbol{a}$,
$\boldsymbol{d} = \boldsymbol{1}$, the inequality \eqref{Cor21}
holds for $x$, satisfying the condition
\begin{equation}\label{example1}
x \leq \dfrac{a_{1}}{(a_{1}-1)}\sum\limits_{i=1}^{n}\ln a_{i} =
\frac{2a_{1}\ln D}{a_{1}-1},  \qquad
D^{2} = \prod\limits_{i=1}^{n}a_{i} > 1.
\end{equation}

In order to apply Theorem 1 (i.e. the condition \eqref{prop1ad})
note that if $\boldsymbol{a}$ - monotone decreasing vector, then
the function $f(i,\boldsymbol{a},\boldsymbol{1}) = 1-1/a_{i}$
also monotonically decreases in $i$, and therefore
$k_{1}(\boldsymbol{a},\boldsymbol{1}) = 1$. Using formulas
\eqref{prop1am1} and \eqref{defk1} we get
$$
\begin{gathered}
T(1,\boldsymbol{a},\boldsymbol{1}) = 1 - D^{-2/n}.
\end{gathered}
$$
Therefore if $D > 1$, then the condition \eqref{prop1ad}
has the form
\begin{equation}\label{example11}
x \leq \frac{2\ln D}{T(1,\boldsymbol{a},\boldsymbol{1})} =
\frac{2\ln D}{1 - D^{-2/n}}= nf\left(D^{2/n}\right),
\end{equation}
where
$$
\begin{gathered}
f(z) = \frac{z\ln z}{z-1}, \qquad f'(z) > 0,\qquad z > 1.
\end{gathered}
$$
The function $f(z)$ monotonically increases from $f(1) = 1$ up to
$f(\infty) = \infty$. In particular, if $D \geq 1$, then the
inequality \eqref{example11} holds for $x \leq n$. The condition
\eqref{example11} may be much better than the condition
\eqref{example1}.

Theorem 2 (i.e. the condition \eqref{stren22}) may give result
better than \eqref{example11}, and worse than it as well
(depending on the vector $\boldsymbol{a}$).

Next example gives the less obvious inequality, opposite to
\eqref{Cor21}.

2. Let $\boldsymbol{a} = \boldsymbol{1}$,
$b_{1} > 1$, $b_{1} > b_{2} \geq \ldots \geq b_{n}$ and
$b_{1}b_{2} < 1$. Theorem 1 from \cite{Bak95} is not applicable
here, since the condition \eqref{Bak2} is not fulfilled.
The function $f(i,\boldsymbol{1},\boldsymbol{b}) = 1/b_{i}-1$
monotonically increases in $i$ and
$\max\limits_{i}f(i,\boldsymbol{1},\boldsymbol{b}) = 1/b_{n}-1$.
Therefore Theorem 1 does not improve the estimate \eqref{prop1a}.
If $\sum\limits_{i=1}^{n}\ln b_{i} < 0$, then from \eqref{prop1a}
we get
\begin{equation}\label{Ex2}
\begin{gathered}
\mathbf{P}\left(\sum\limits_{i=1}^{n}\xi_{i}^{2}< x\right) \leq
\mathbf{P}\left(\sum\limits_{i=1}^{n}b_{i}\xi_{i}^{2}< x\right),
\qquad x \leq \frac{b_{n}}{(1-b_{n})}
\sum\limits_{i=1}^{n}\ln \frac{1}{b_{i}}.
\end{gathered}
\end{equation}

In order to apply Theorem 2 (i.e. the condition \eqref{stren22})
notice that $n_{1} = n_{1}(\boldsymbol{a},\boldsymbol{b}) = 2$,
$d = 1/b_{2}-1$. Therefore we get
\begin{equation}\label{Ex2a}
\begin{gathered}
\mathbf{P}\left(\sum\limits_{i=1}^{n}\xi_{i}^{2}< x\right) \leq
\mathbf{P}\left(\sum\limits_{i=1}^{n}b_{i}\xi_{i}^{2}< x\right),
\qquad x \leq \frac{b_{2}}{(1-b_{2})}
\left[\ln \frac{1}{b_{1}} + (n-1)\ln \frac{1}{b_{2}}\right].
\end{gathered}
\end{equation}
The condition \eqref{Ex2a} may be much broader than \eqref{Ex2}.

The author thanks the reviewer for constructive critical remarks,
which helped to improve the paper material presentation.

%\newpage

%\end{thebibliography}

\vspace{5mm}

\begin{flushleft}
{\small {\it Burnashev Marat Valievich} \\
Kharkevich Institute for Information Transmission Problems, \\
Russian Academy of Sciences, Moscow\\
 {\tt burn@iitp.ru}}
% {\small {\it Бурнашев Марат Валиевич} \\
%Институт проблем передачи информации \\
%им. А.А. Харкевича РАН \\
% {\tt burn@iitp.ru}}
\end{flushleft}%


\vspace{5mm}

\begin{center}{\large REFERENCES} \end{center}
\begin{enumerate}
%\begin{thebibliography}{10}
\bibitem{Bak95}
{\it Bakirov N. K.} Comparison Theorems for Distribution
Functions of Quadratic Forms of Gaussian Vectors //
Theory of Probability and Its Applications. 1995. V. 40. no. 2.
P. 404--412.
%Теоремы сравнения для функций распределения
%квадратичных форм от гауссовских величин // Теория вероятн. и ее
%примен. 1995. Т. 40. № 2. С. 404--412.
\bibitem{Wald}
{\it Wald A.} Statistical Decision Functions. New York:
Wiley, 1950.
%А. Вальд, Статистические решающие функции, в сб.
%«Позиционные игры», 300—-522, М., изд-во «Наука», 1967.
\bibitem{Leman}
{\it Lehmann E. L.} Testing of Statistical Hypotheses.
New York: Wiley, 1959.
%{\it Леман Э.} Проверка статистических гипотез. М.: Наука, 1979.
\bibitem{Bur79}
{\it Burnashev M. V.}
On the Minimax Detection of an Inaccurately Known Signal in a
White Gaussian Noise Background //
Theory of Probability and Its Applications. 1979. V. 24. no. 1.
P. 106--118.
%О минимаксном обнаружении неточно известного
%сигнала на фоне белого гауссовского шума // Теория вероятн. и ее
%примен. 1979. Т. 24. № 1. С. 106--118.
\bibitem{ZhangPoor11}
{\it Zhang W., Poor H.V.} On Minimax Robust Detection of Stationary
Gaussian Signals in White Gaussian Noise // IEEE Trans. Inform.
Theory. 2011. V. 57. № 6. P. 3915--3924.
\bibitem{Bur17}
{\it Burnashev M. V.} On Detection of Gaussian Stochastic
Sequences // Problems of Information Transmission. 2017 (in print).
%Об обнаружении гауссовских стохастических
%последовательностей // Пробл. передачи информ. 2017.
\bibitem{Ponom85}
{\it Ponomarenko L. S.} On Estimating Distributions of
Normalized Quadratic Forms of Normally Distributed Random
Variables // Theory of Probability and Its \\
Applications. 1985. V. 30. no. 3. P. 545--549.
%Об оценивании распределений нормированных
%квадратичных форм от нормально распределенных случайных величин //
%Теория вероятн. и ее примен. 1985. Т. 30. № 3. С. 545--549.
\end{enumerate}
\end{document}